\documentclass{ws-mplb}
\usepackage{amsfonts,amssymb,amsmath}
\usepackage{graphicx}
\usepackage[super,sort,compress]{cite} 
\usepackage{subfigure}
\begin{document}

\markboth{Yong Zhao, Yao Wu, Zhenhua Chai and Baochang Shi}{A}

%
\catchline{}{}{}{}{}
%

\title{Lattice Boltzmann simulations of melting in a rectangular cavity heated locally from below at high Rayleigh number}

\author{\footnotesize Yong Zhao, Yao Wu, Zhenhua Chai and Baochang Shi \footnote{Corresponding author.}}

\address{Hubei Key Laboratory of Engineering Modeling and Scientific Computing,\\
 School of Mathematics and Statistics, Huazhong University of Science and Technology, Wuhan 430074, China\\
 shibc@hust.edu.cn}

\maketitle

\begin{history}
\received{(Day Month Year)}
\revised{(Day Month Year)}
\end{history}

\begin{abstract}

This work presented a block triple-relaxation-time (B-TriRT) lattice Boltzmann model for simulating melting in a rectangular cavity heated from below at high Rayleigh ($Ra$) number ($Ra=10^8$). The test of benchmark problem show that present B-TriRT can dramatically reduce the numerical diffusion across the phase interface. In addition, the influences of location of the heated region are investigated. The results indicate that the location of heated region plays an essential role in melting rate and the full melting occur earliest when the heated region is located at the middle region.

\end{abstract}

\keywords{Lattice Boltzmann method; solid-liquid phase change; high Rayleigh number.}

\section{Introduction}

Solid-liquid phase change problem has attracted significant interest due to its universality, and usually encounter in applied engineering fields, including latent heat thermal energy storage~\cite{zhang2018latent}, battery thermal management\cite{malik2016review}, building walls\cite{konuklu2015review} and so on. Melting in a rectangular enclosure has been investigated by a number of analytical, numerical and experimental studies. However, the melting of phase change material (PCM) heated from below is rarely reported. It is well known that when heated from below, the heat transfer structure is similar to the classical Rayleigh-B$\acute{\mathrm{e}}$nard convection, which may lead to turbulent thermal convection at high Rayleigh number.~\cite{feng2015numerical}. 

An in-depth understanding of the complex transport mechanism in thermal convective flows requires powerful experimental and computational tools. Lattice Boltzmann method (LBM), as a promising numerical method, has been developed into a  powerful numerical tool for computational fluid dynamics and heat transfer in the past thirty years. Although many LB models have been proposed for the solid-liquid phase change problem, the most widely used models are the single-relaxation-time (SRT) model. As well know, the SRT model is numerically
most efficient, but it is usually unstable for high Rayleigh number problem. In addition, as reported by Huang~\cite{huang2015phase}, there is a numerical diffusion across the phase interface for SRT model. However, Huang adopted the multiple-relaxation-time (MRT) model to reduce the numerical diffusion, it is not convenient to determine the free relaxation parameters in MRT model. Therefore, we present a block triple-relaxation-time (B-TriRT) model to solve the phase change problem, which has less relaxation parameters to be determined than MRT model, and can also reduce the numerical diffusion. 


\section{The block triple-relaxation-time lattice Boltzmann model}

For solid-liquid phase change problem, the energy conservation equation with incompressible flow can be given as
\begin{equation}
\frac{\partial H}{\partial t}+\nabla \cdot (C_p T\mathbf{u})=\frac{1}{\rho_0}\nabla\cdot(\kappa\nabla T),\\
\label{eq1}
\end{equation}
where $C_p, \kappa, \rho_0$ are the specific heat,  thermal conductivity and referenced density, $H=C_pT+f_lL$ is the total enthalpy with $f_l, L$  being liquid fraction and latent heat.

To attain a high stability, the block triple-relaxation-time lattice Boltzmann model~\cite{zhao2019block} is employed. In B-TriRT model, the evolution of distribution function is given as
\begin{equation}
\begin{aligned}\label{eq2}
f_i(\textbf{x}+\textbf{c}_i\Delta t,t+\Delta t)=& f_i(\textbf{x},t)-k_0f_i^{neq}(\textbf{x},t)-(k_1-k_0)\frac{\omega_i\textbf{c}_i\cdot \textbf{M}_1^{neq}(\textbf{x},t)}{c_s^2}\\
&-(k_2-k_0)\frac{\omega_i(\textbf{c}_i\textbf{c}_i-c_s^2\textbf{I}):\textbf{M}_2^{neq}(\textbf{x},t)}{2c_s^4},
\end{aligned}
\end{equation}
where $k_0,~k_1,~k_2$ are dimensionless relaxation parameters, $\textbf{M}_1^{neq}(\textbf{x},t)=\sum_j\textbf{c}_j f^{neq}_j, ~\textbf{M}_2^{neq}(\textbf{x},t)=\sum_j \textbf{c}_j \textbf{c}_j f^{neq}_j$ are the first-order and second-order moment of non-equilibrium distribution function $f^{neq}_j=f_j-f^{eq}_j$.

In addition, the equilibrium distribution function is defined as
\begin{equation}
g^{eq}_i=
\begin{cases}
H-C_pT+\omega_iC_pT,&i=0,\\
\omega_iC_p T\left[1+\frac{\textbf{c}_i\cdot\textbf{u}}{c^2_{s}}\right],&i\neq0,
\end{cases}
\label{eq4}
\end{equation}
where $\omega_i$ is the weight coefficient. For the two-dimensional case considered here, the popular D2Q9 model is employed, then the weight coefficient can be give by $\omega_0=4/9,~\omega_i=1/9~(i=1-4),~\omega_i=1/36~(i=5-8)$, respectively.  In order to recover the correct energy equation Eq.~(\ref{eq1}), the dimensionless relaxation time $k_1$ must satisfy
\begin{equation}
\frac{1}{k_1}=\frac{\kappa}{\rho_0 C_p c_s^2\Delta t}+0.5=\frac{\alpha}{c_s^2\Delta t}+0.5,
\label{eq5}
\end{equation}
where $\alpha$ is the thermal diffusivity. As as pointed out in Ref.~\cite{huang2015phase}, the traditional total enthalpy-based thermal LBM is usually accompanied by the numerical diffusion across the phase interface. To address such drawback, Huang et al. proposed an improved MRT model to reduce the numerical diffusion. We would like to point out that the present B-TriRT model can be written as an MRT form. Furthermore, based on the analysis performed by Huang, we also derive a relational expression to reduce the numerical diffusion as
\begin{equation}
k_1+k_2=2.
\label{eq6}
\end{equation}
By adopting the relational expression, the solid phase can be precisely kept at the melting temperature and the phase interface can be exactly restricted in one lattice spacing, i.e., the numerical diffusion across phase interface can be thoroughly eliminated, which will be confirmed in the next Section.

\section{Numerical Results}

\subsection{One-phase melting by conduction}
We first consider a benchmark problem to validate the present B-TriRT model in reducing numerical diffusion across the phase interface. The problem of one-phase melting  by conduction is described as follows, initially, the phase change material is solid uniformly and the temperature is set to be $T_m$. At time $t>0$, a constant temperature $T_h (T_h>T_m)$ is imposed on the left wall (i.e., x=0), then the phase change material begins to melt. For this problem, the analytical solution of temperature is given by \cite{huang2015phase}
\begin{equation}
T(x,t)=
\begin{cases}
T_h-\frac{(T_h-T_m)erf\left(\frac{x}{2\sqrt{\alpha t}}\right)}{erf(\kappa)},&0<x<X_i(t),\\
T_m,&x>X_i(t),
\end{cases}\label{eq7}
\end{equation}
where $erf(x)=\frac{2}{\sqrt{\pi}}\int_0^x e^{-\eta^2}d\eta$ is the error function, and the position of phase interface can be determined as
\begin{equation}
X_i (t)=2\kappa\sqrt{\alpha t}, ~~~~\frac{Ste}{exp(\kappa^2)erf(\kappa)}=\kappa\sqrt{\pi}.
\label{eq8}
\end{equation}

In the present simulations, the dimensionless parameters and thermophysical properties are set as $T_h=1, T_m=0, Ste=0.01, \alpha=1/6$, and the computational parameters of present model are given as $\Delta x=1/100, c=1.0$. In addition, as for the time step $\Delta t$, it is obtained by the relationship $\Delta t=\alpha/[c^2_s(1/k_1-0.5)]$ with $k_1$ being set as $0.1,~0.5,~1.0,~1.5,$ and $~2.0$, respectively. The temperature distributions at $t=25$ of SRT model and present B-TriRT model for different $k_1$ are illustrated in Fig.~1. It can be seen from Fig.~1 (a), for SRT model, the temperature distributions deviate from the analytical solutions when $k_1 \neq 1$. The results show that the solid phase has been activated has been before it starts melting, which can be explained as the numerical diffusion \cite{huang2015phase}. However, for present B-TriRT model, the temperature distributions agree well with analytical solutions, and the temperatures of solid phase stay at $T_m$. The results indicate that the numerical diffusion across the phase interface can be dramatically reduced when employing present B-TriRT model. On the other hand, as discussed in Ref.~\cite{huang2015phase}, the numerical diffusion will also cause the phase interface width exceeding one lattice spacing. To show this clearly, we also captured the liquid fraction distributions at $t=25$ of SRT model and present B-TriRT model for different $k_1$, and plotted the results in Fig.~2. The results show that the phase interface width obtained from SRT model is always larger than  one lattice spacing. However, the phase interface width obtained from present improved B-TriRT model is exactly in one lattice spacing. Based on the above results and discussions, it can be concluded that the present B-TriRT model can be exploited to reduce the numerical diffusion across the phase interface in the solid-liquid phase change problem.

\begin{figure}[ht]
\centering
\subfigure[]{ \label{fig1:a}
\includegraphics[scale=0.25]{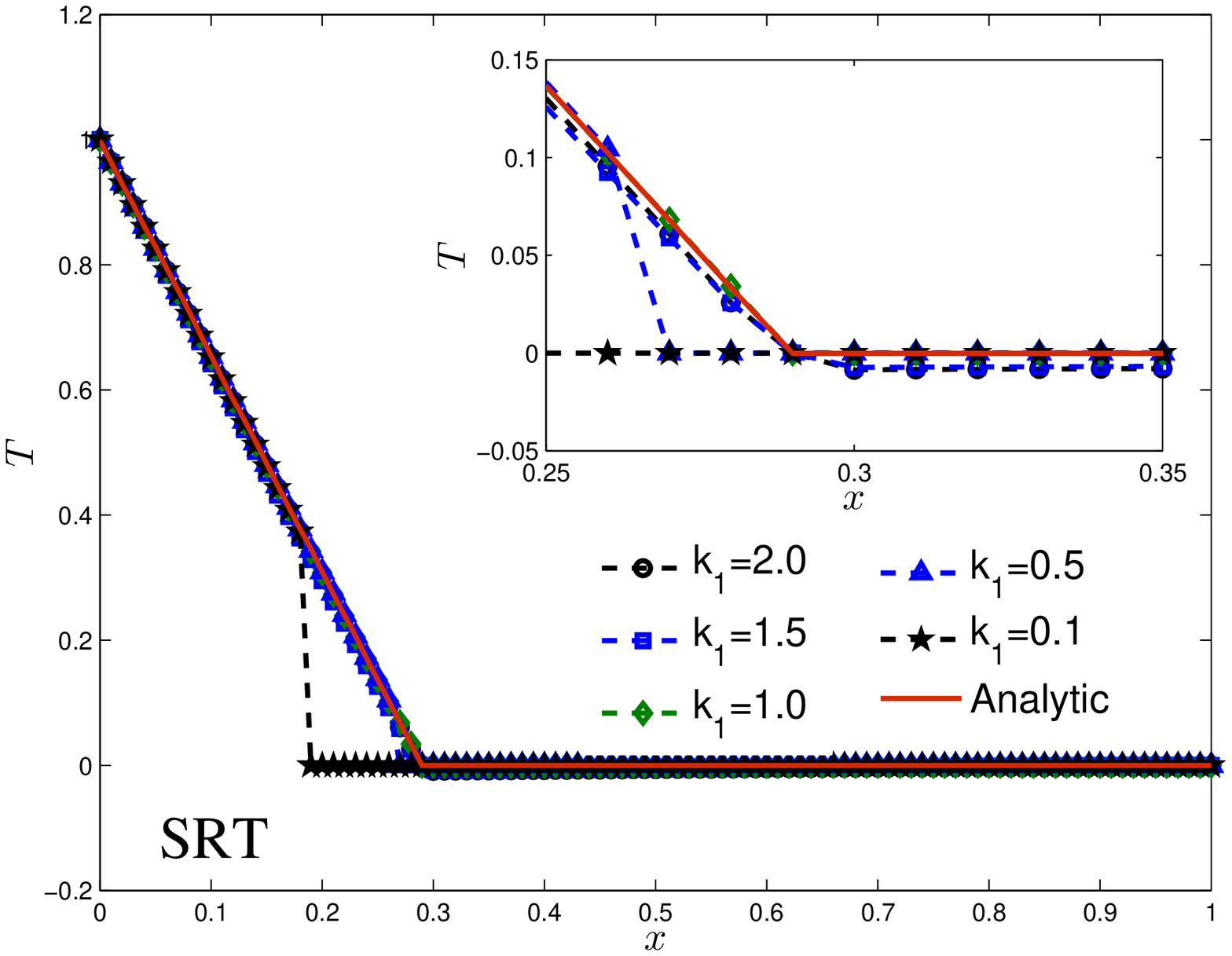}}
\hspace{20pt}
\subfigure[]{ \label{fig1:b}
\includegraphics[scale=0.25]{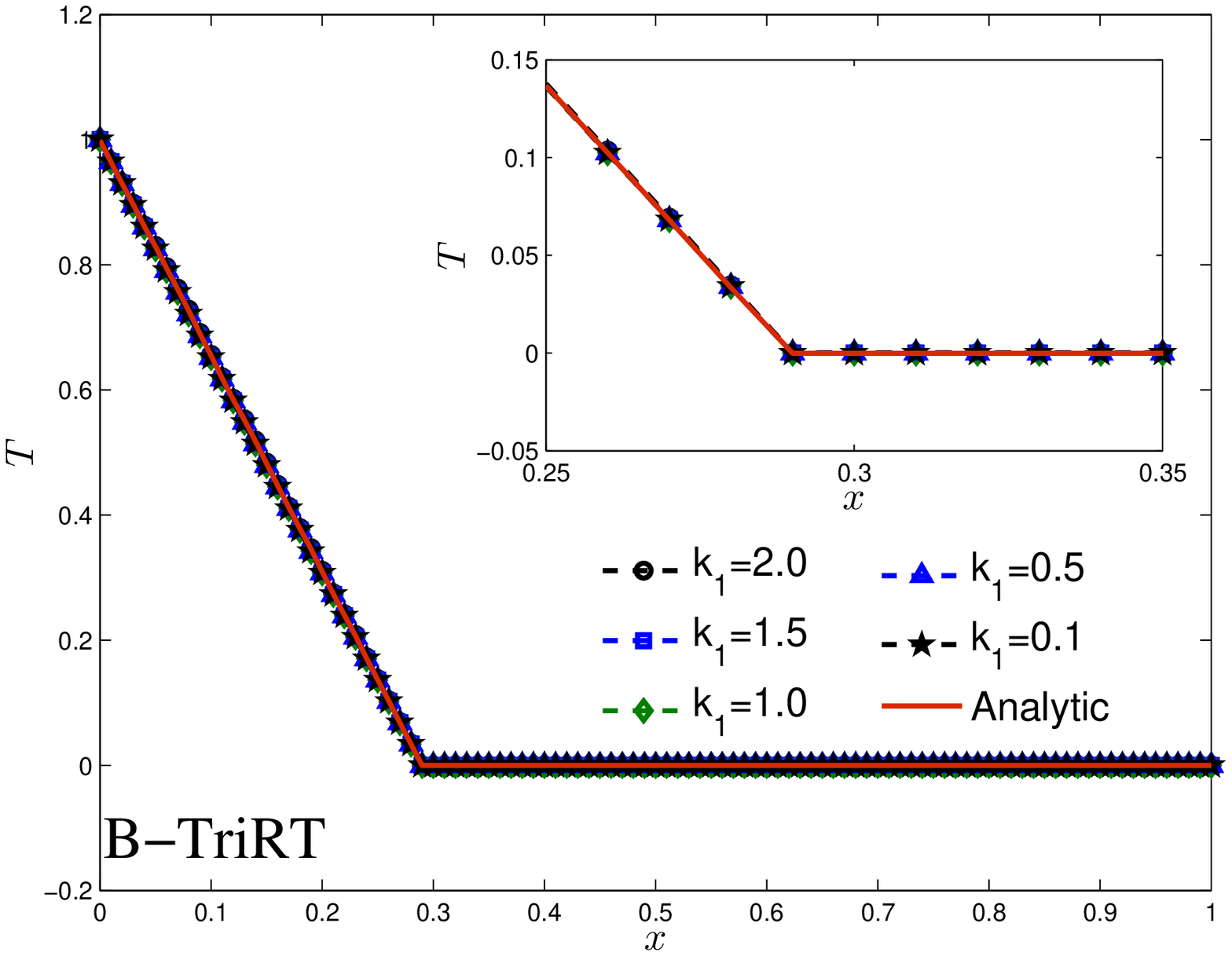}}

\caption{Comparisons of temperature distributions between SRT model (a) and B-TriRT model (b) for different $k_1$.} \label{Fig1}
\end{figure}

\begin{figure}[ht]
\centering
\subfigure[]{ \label{fig2:a}
\includegraphics[scale=0.25]{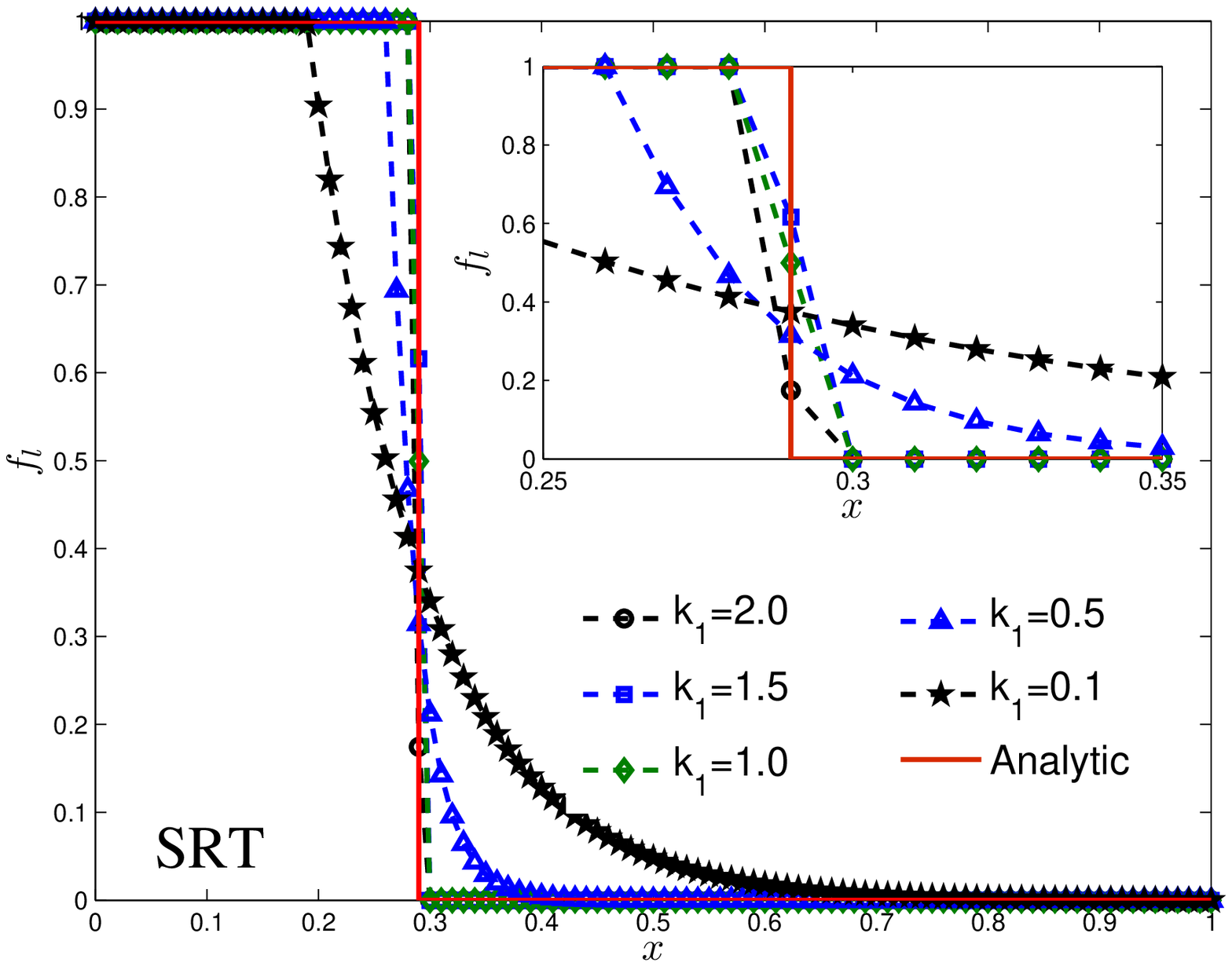}}
\hspace{20pt}
\subfigure[]{ \label{fig2:b}
\includegraphics[scale=0.25]{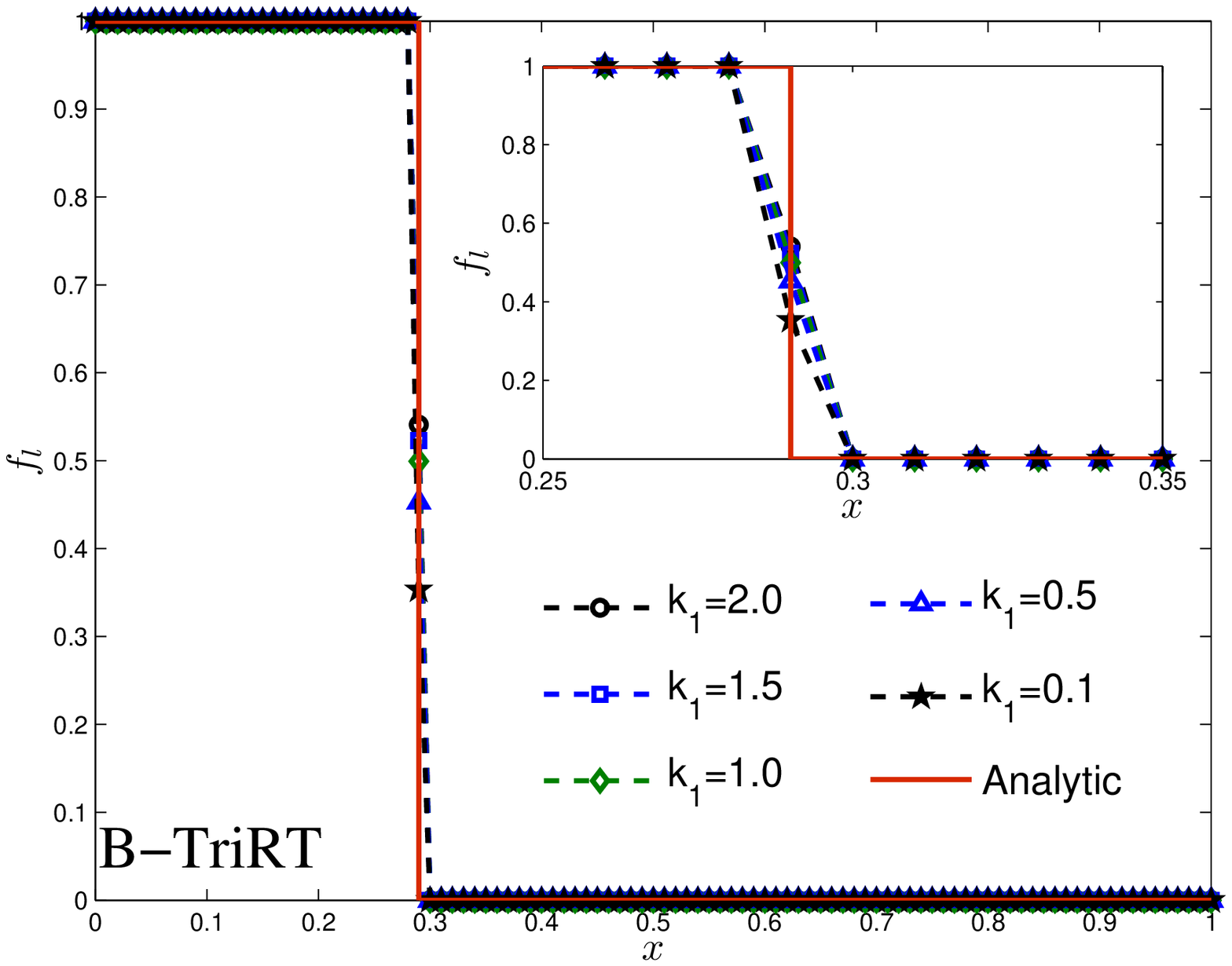}}

\caption{Comparisons of liquid fraction distributions between SRT model (a) and B-TriRT model (b) for different $k_1$.} \label{Fig2}
\end{figure}

\subsection{Melting in a cavity heated locally from below}

\begin{figure}[ht]
\centering
\subfigure[]{ \label{fig3:a}
\includegraphics[scale=0.25]{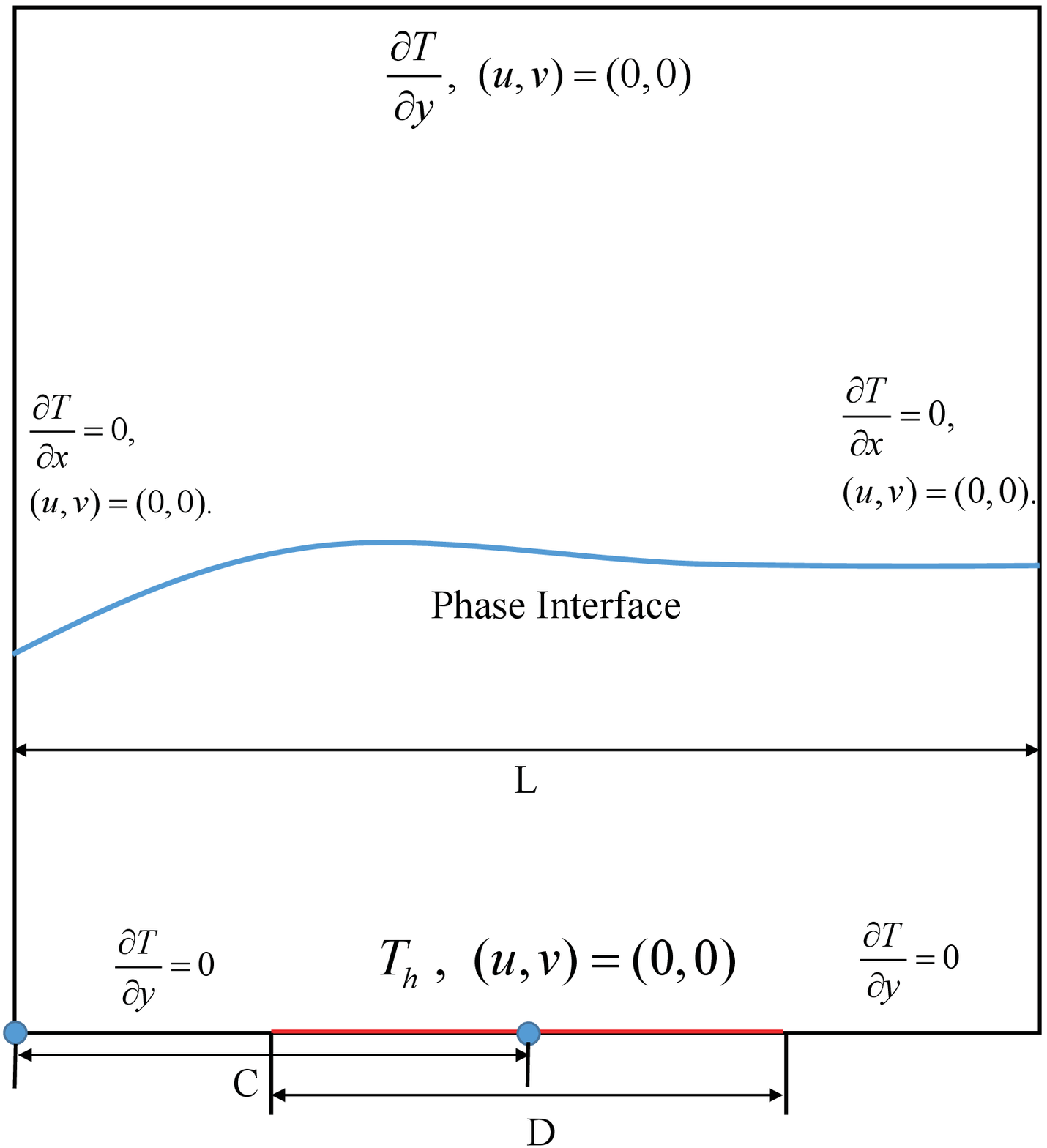}}
\hspace{20pt}
\subfigure[]{ \label{fig3:b}
\includegraphics[scale=0.25]{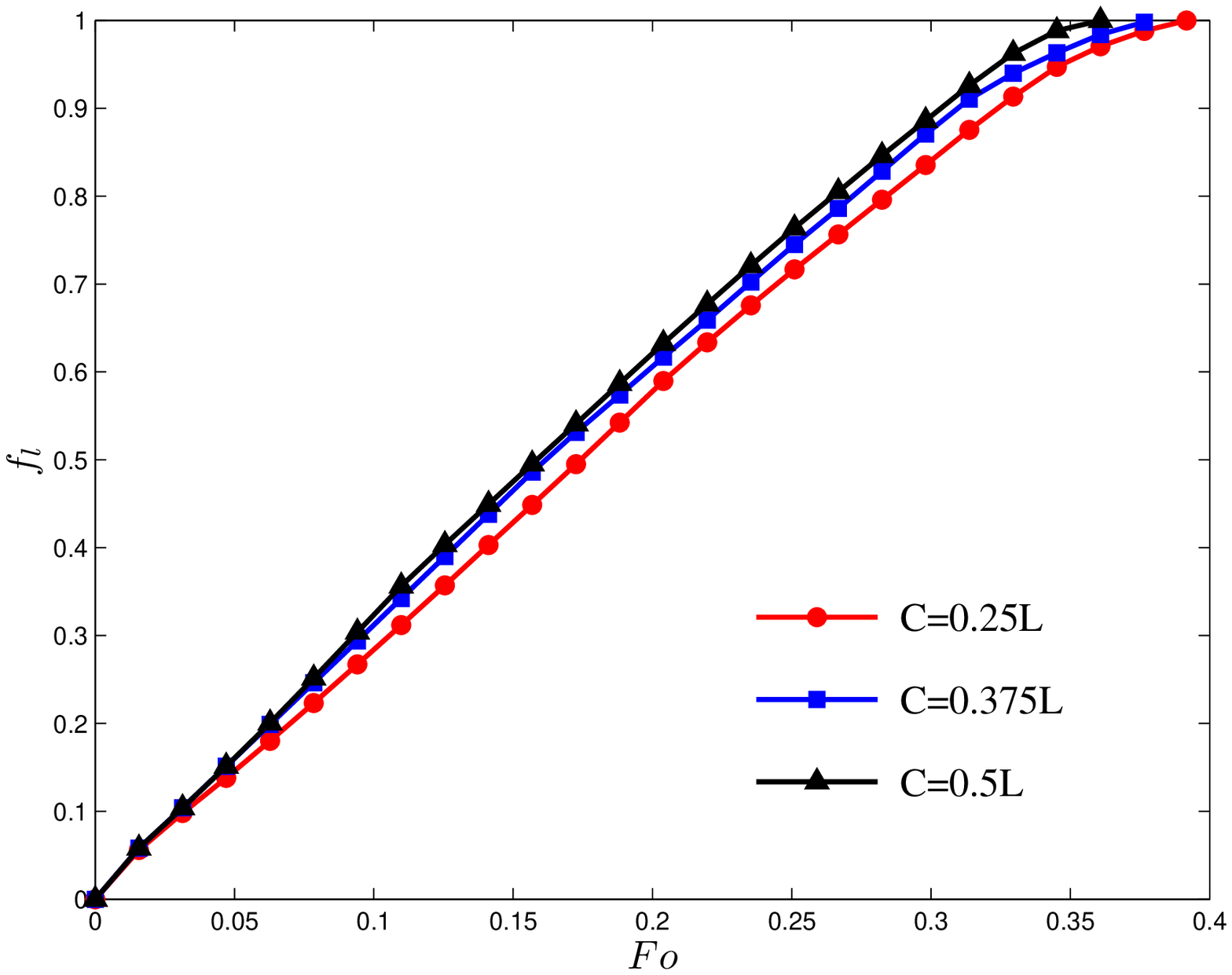}}

\caption{Comparisons of liquid fraction distributions between SRT model (a) and B-TriRT model (b) for different $k_1$.} \label{Fig3}
\end{figure}

We now turn to investigated the melting in a cavity heated locally form below. As shown in Fig.~\ref{fig3:a}, The problem consists of a cubical cavity with length $L=1.0$, filled with PCM which is at the initial temperature $T_m$, and the cavity is heated locally with temperature $T_h$ from below. The width of the heated region is set to be $D=0.5L$, and the distance from left wall to the center of heated region (C) is is varied in the range of $0.25\leqslant C \leqslant 0.5$. For the walls of cavity, the non-slip adiabatic boundary conditions (i.e., $\partial T/\partial \mathbf{n}=0, (u,v)=(0,0)$, where $\mathbf{n}$ is the normal vector of walls) are applied.

\begin{figure}[ht]
\centering
\subfigure[]{ \label{fig4:a}
\includegraphics[scale=0.12]{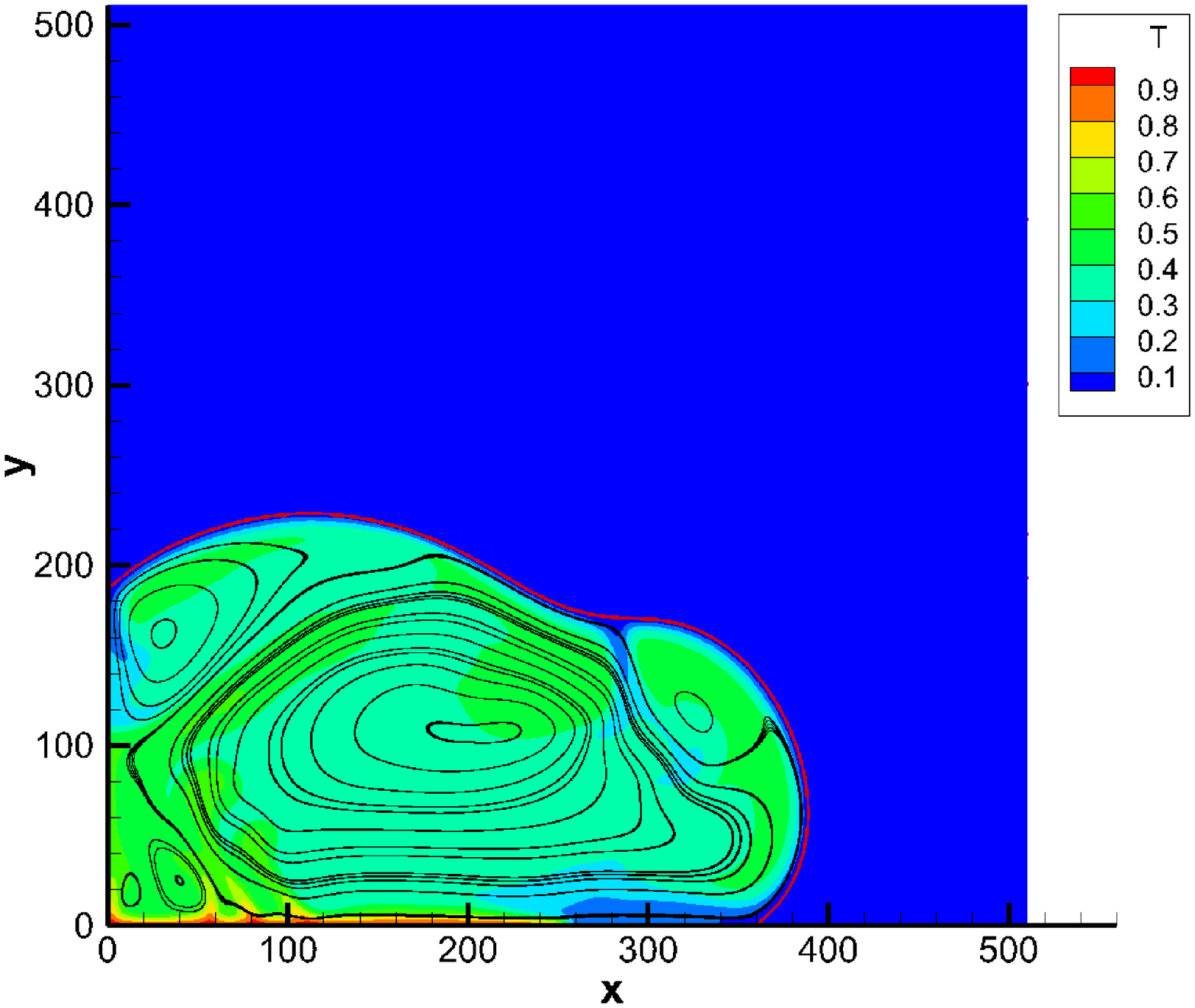}}
\subfigure[]{ \label{fig4:b}
\includegraphics[scale=0.12]{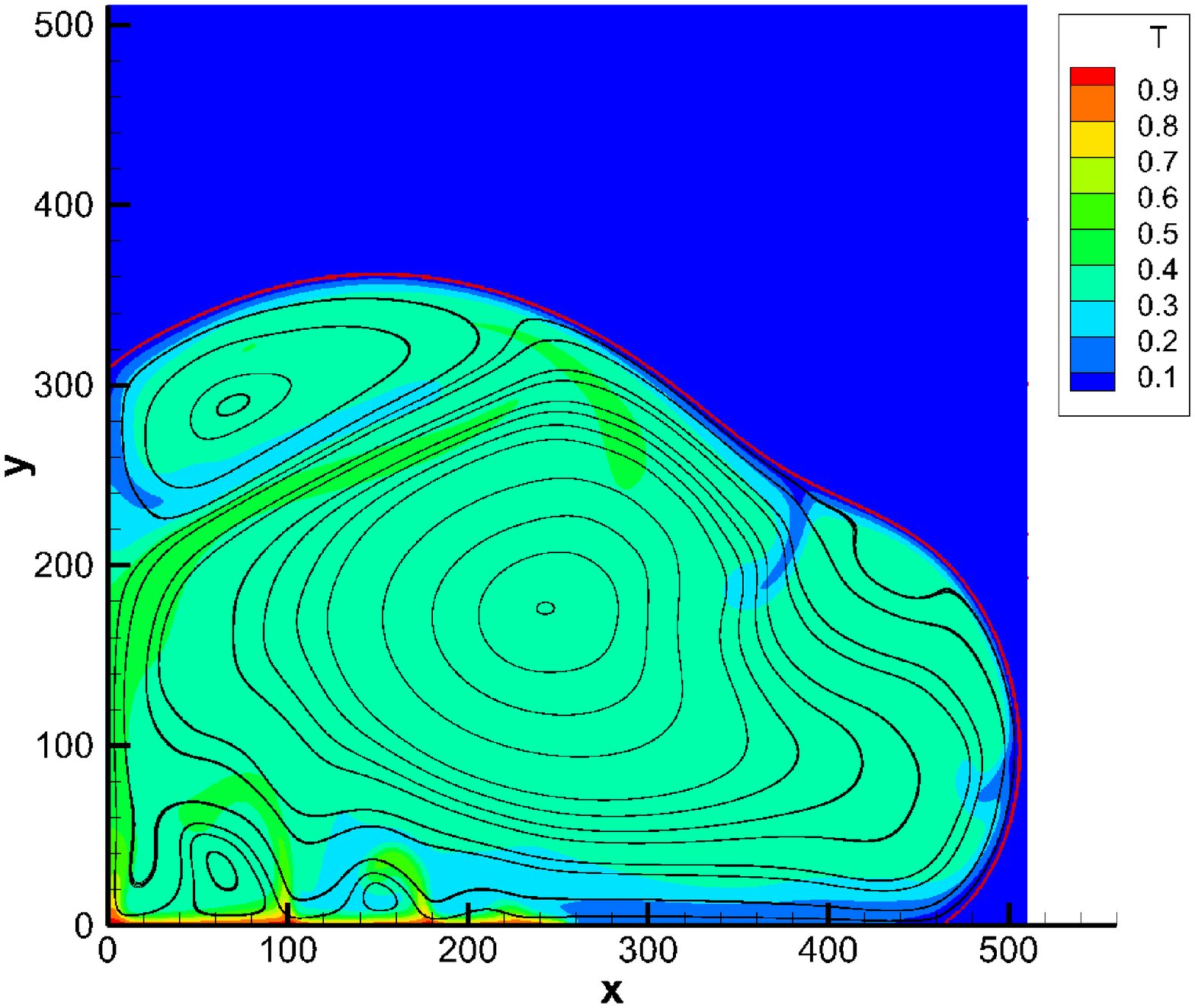}}
\subfigure[]{ \label{fig4:c}
\includegraphics[scale=0.12]{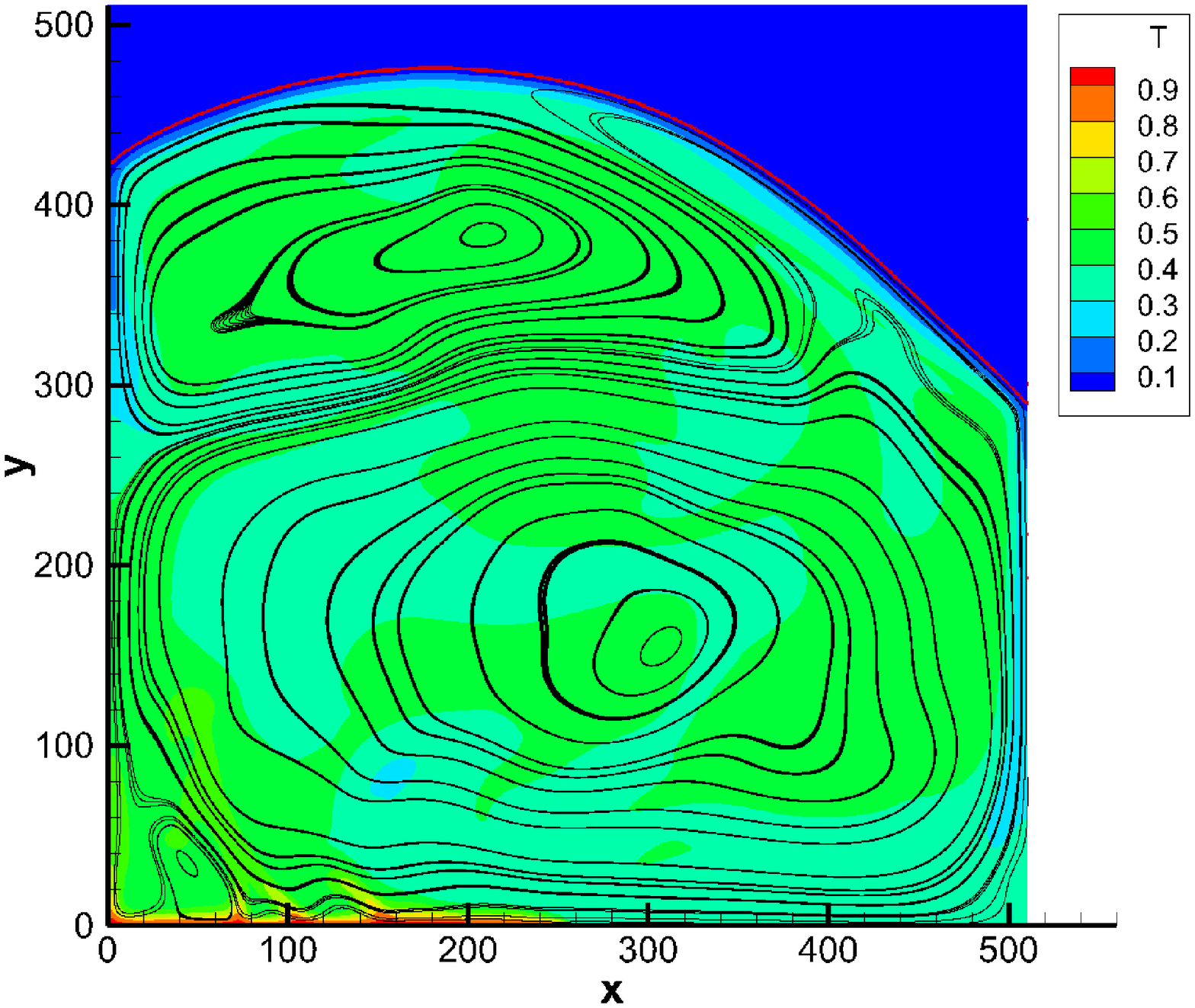}}
\subfigure[]{ \label{fig4:d}
\includegraphics[scale=0.12]{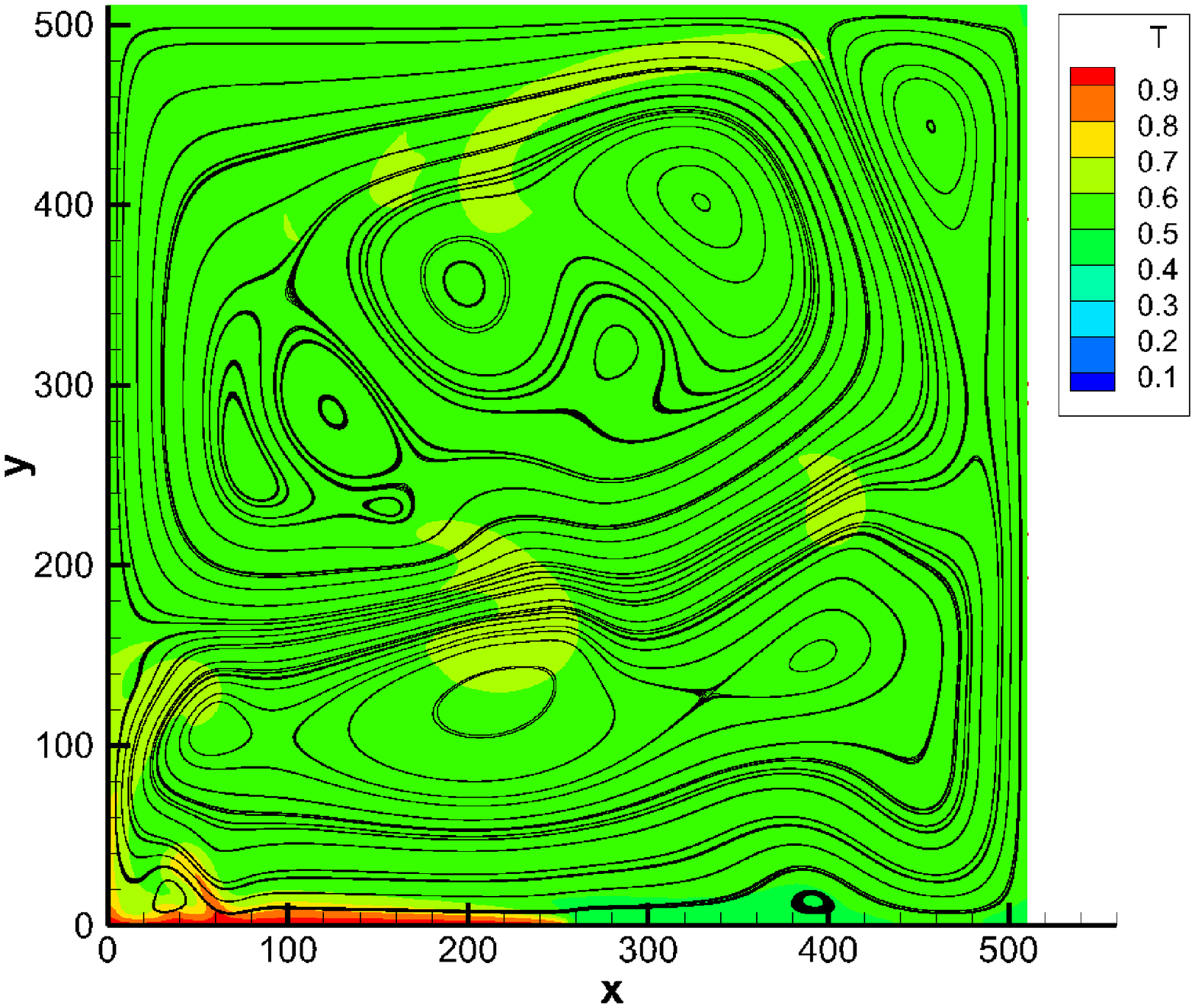}}

\subfigure[]{ \label{fig5:a}
\includegraphics[scale=0.12]{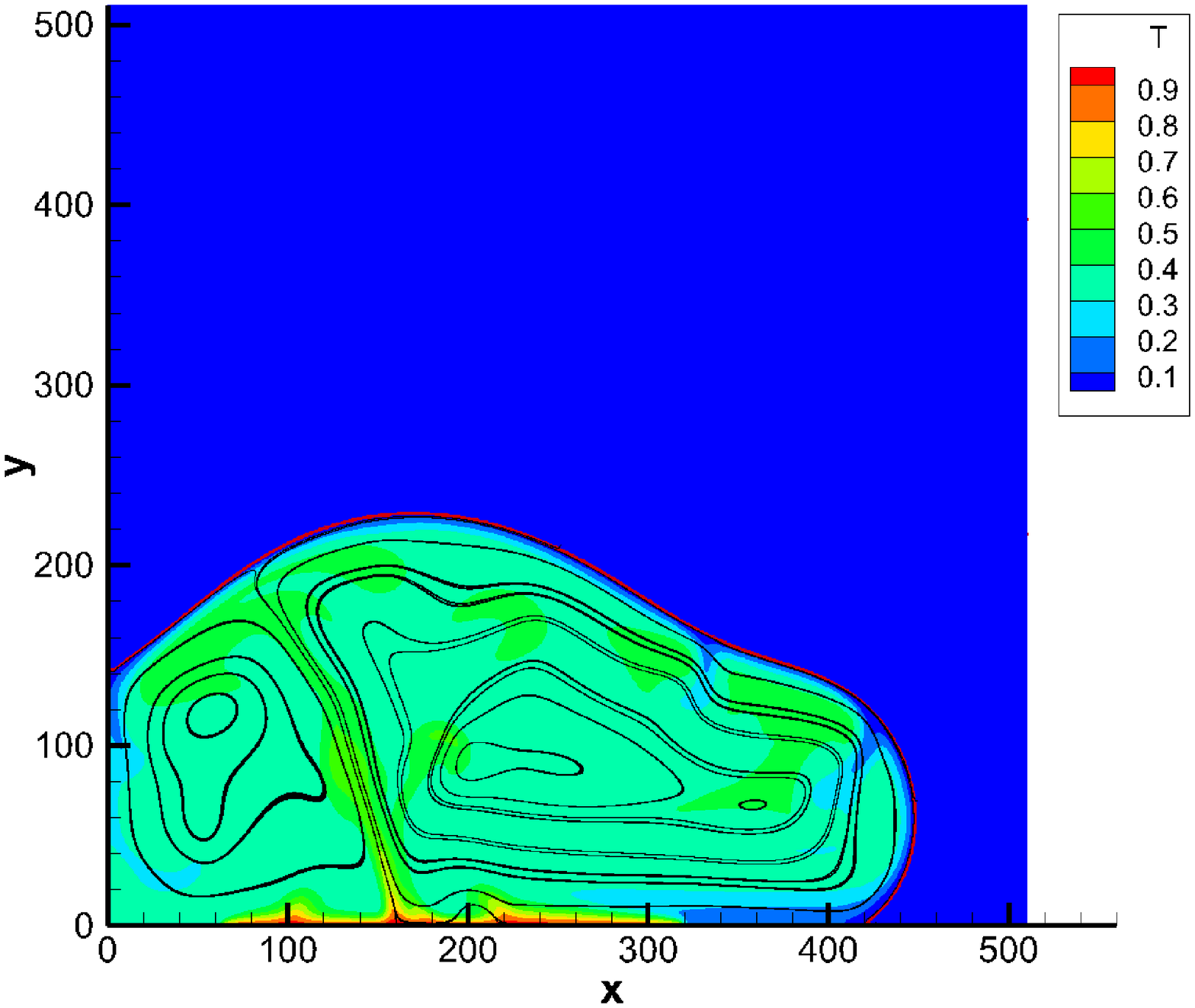}}
\subfigure[]{ \label{fig5:b}
\includegraphics[scale=0.12]{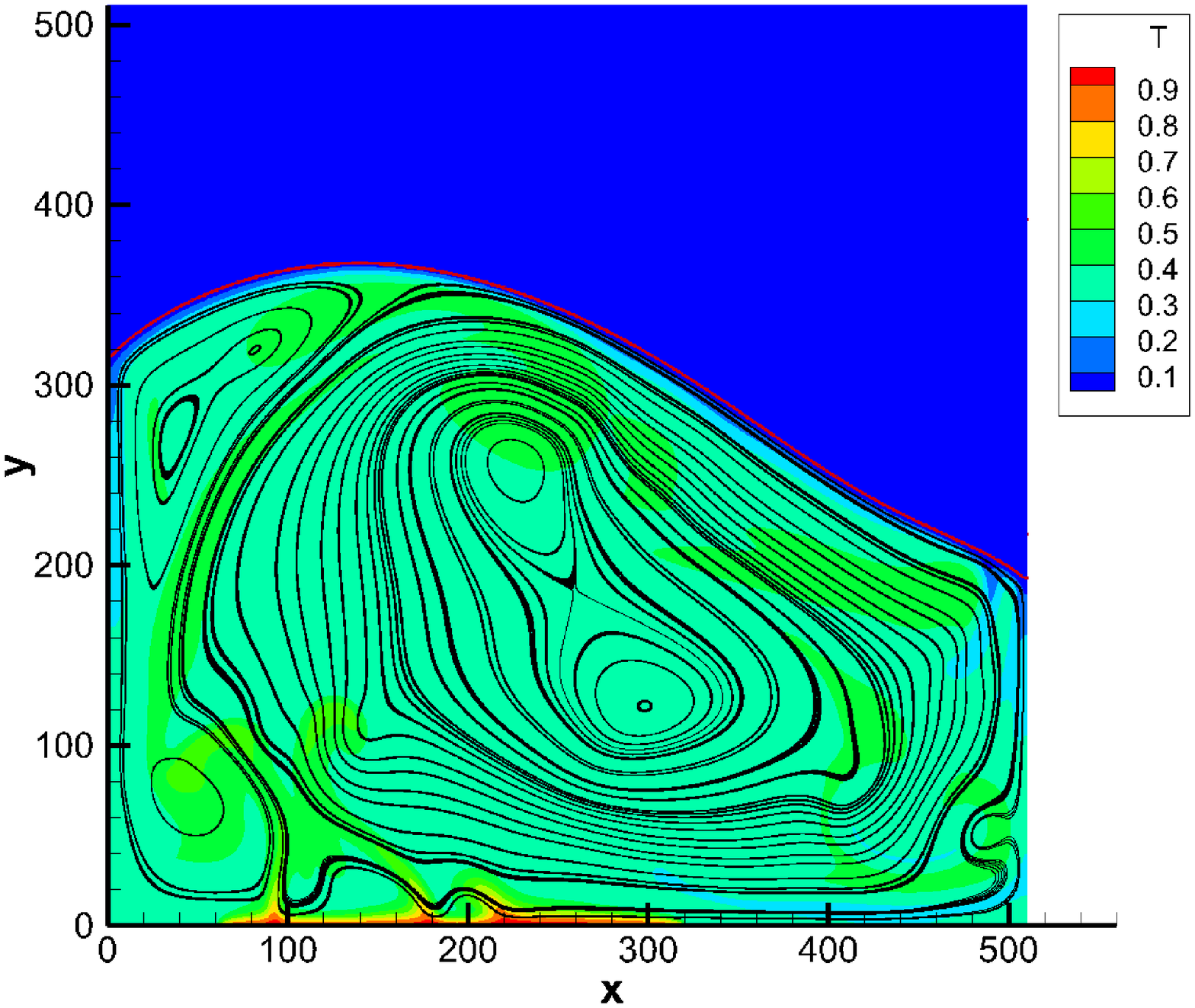}}
\subfigure[]{ \label{fig5:c}
\includegraphics[scale=0.12]{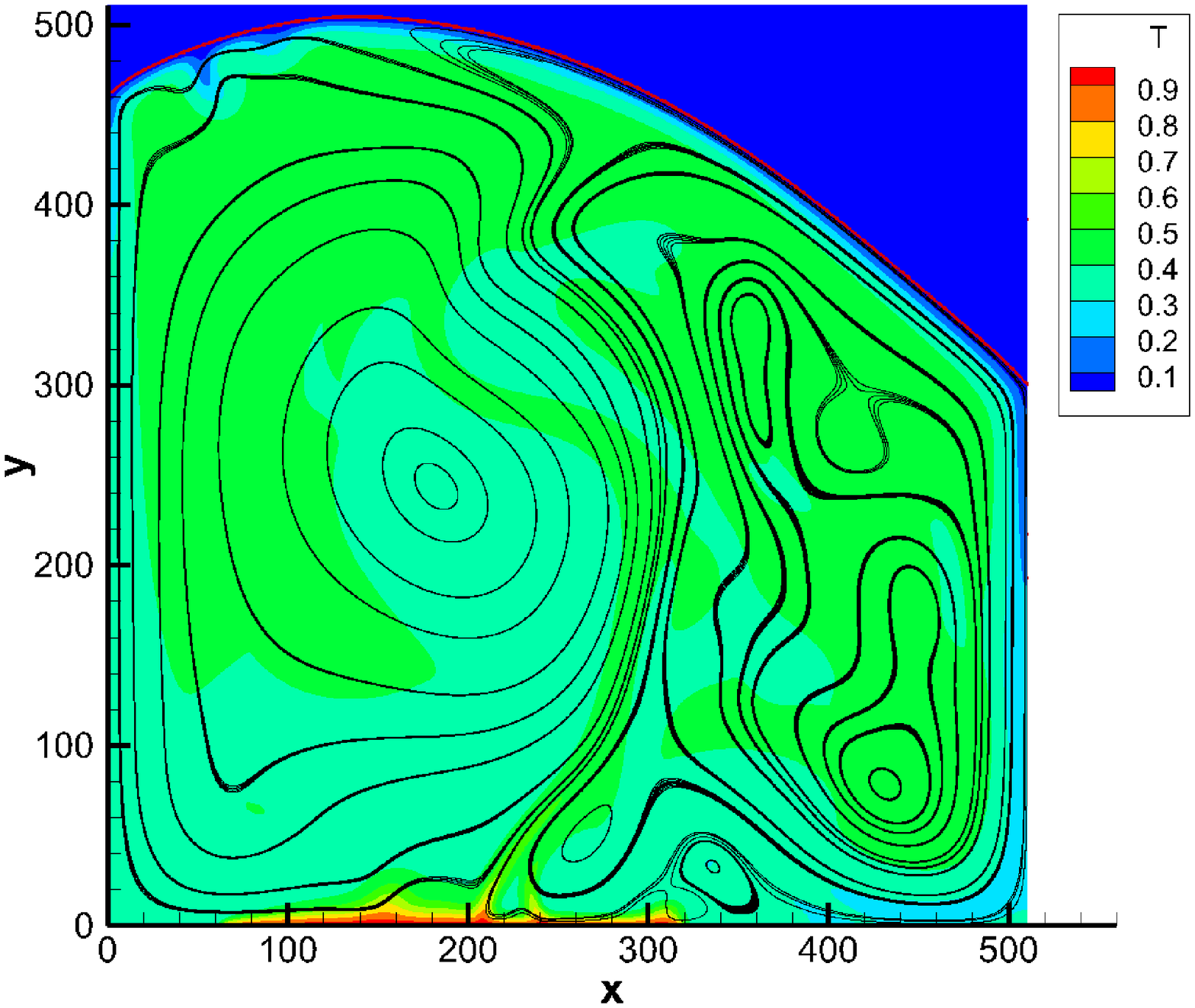}}
\subfigure[]{ \label{fig5:d}
\includegraphics[scale=0.12]{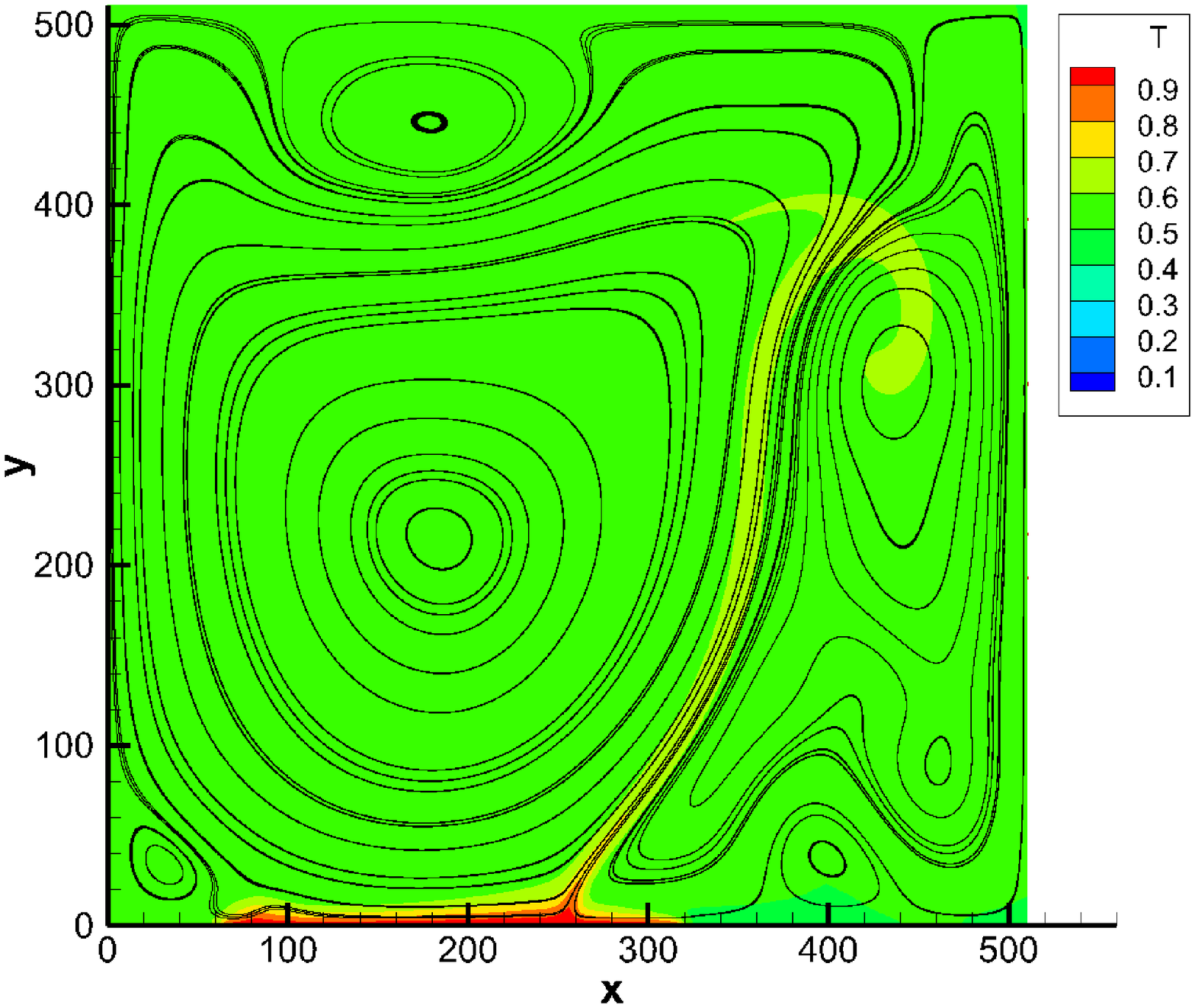}}

\subfigure[]{ \label{fig6:a}
\includegraphics[scale=0.12]{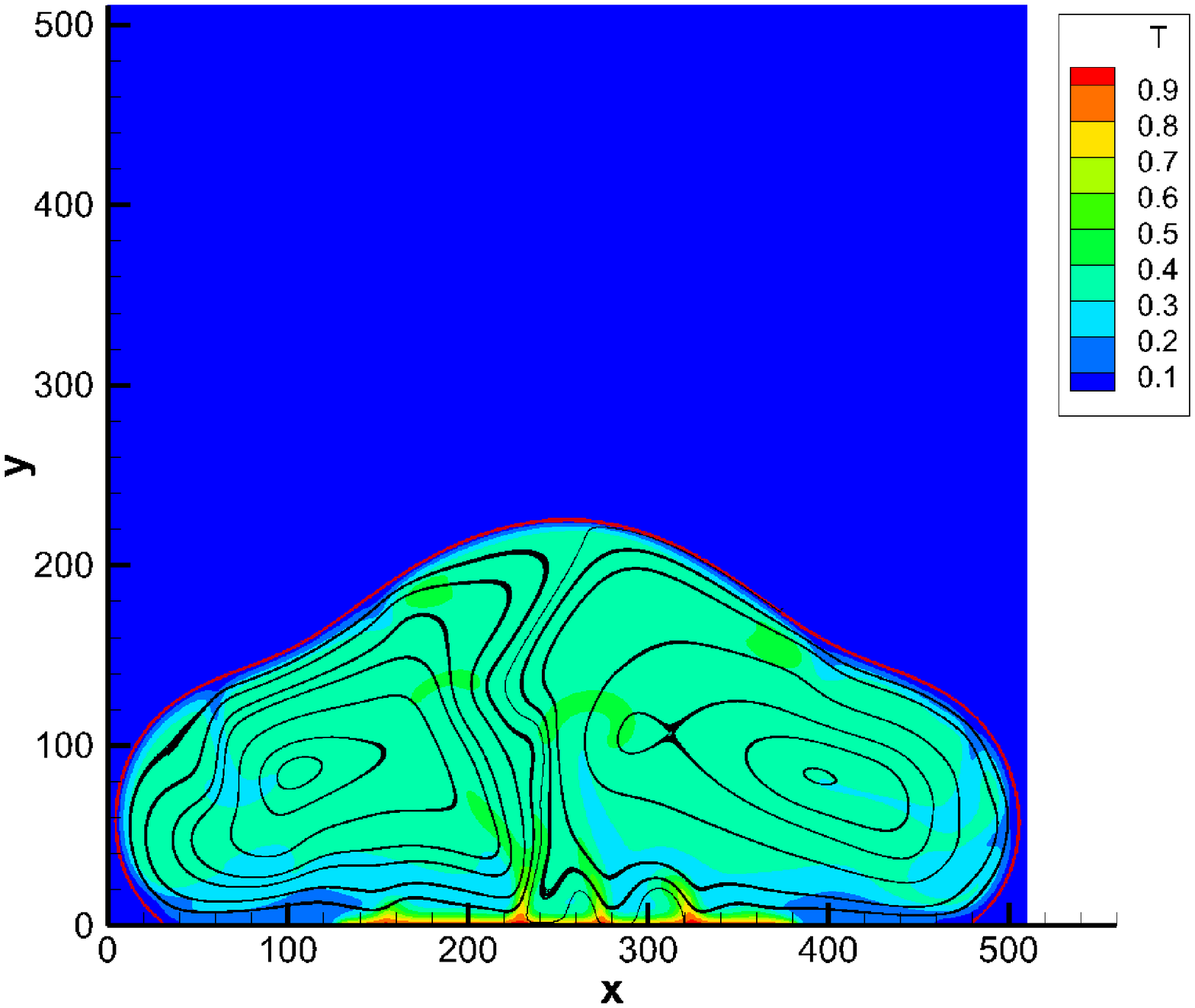}}
\subfigure[]{ \label{fig6:b}
\includegraphics[scale=0.12]{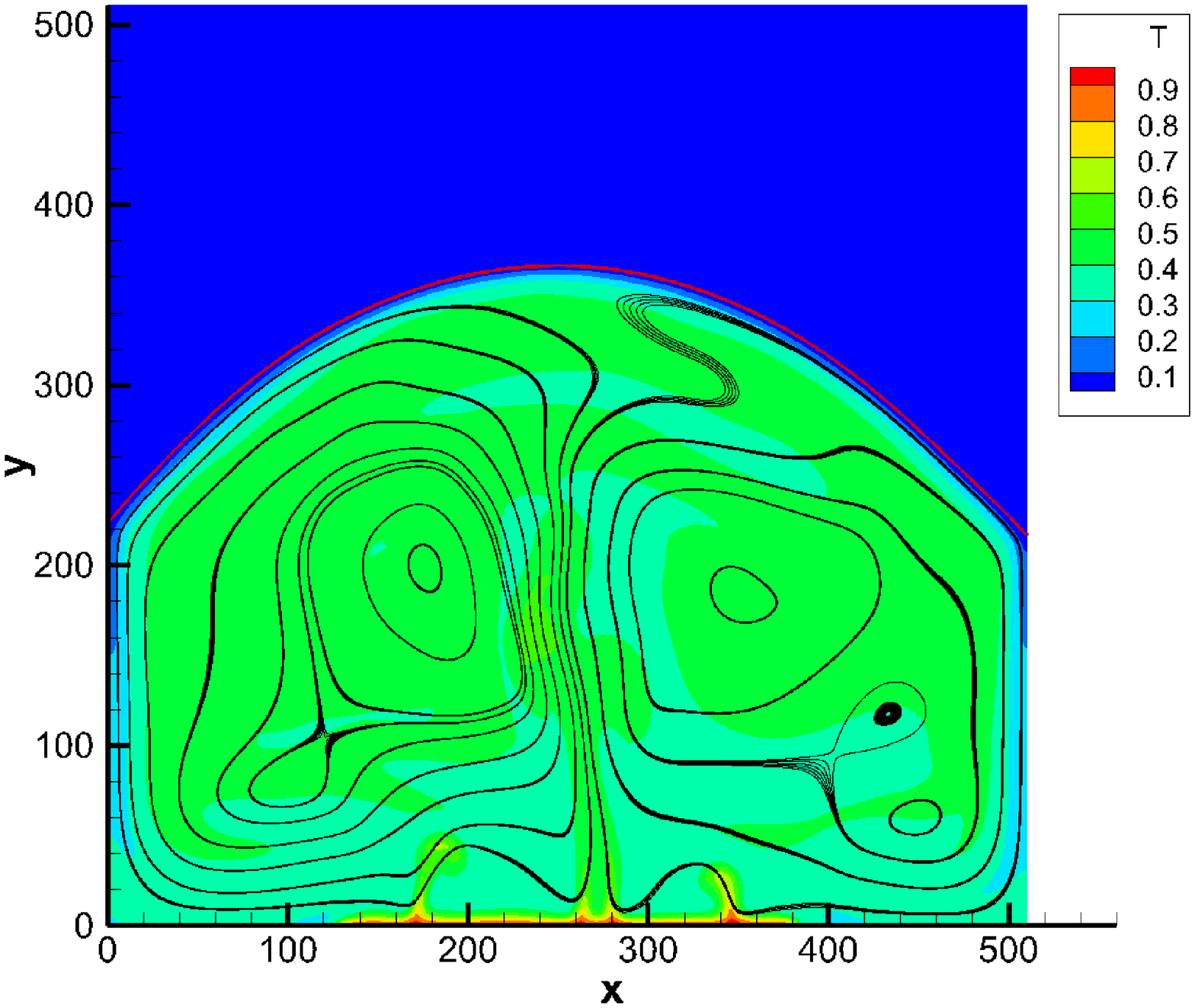}}
\subfigure[]{ \label{fig6:c}
\includegraphics[scale=0.12]{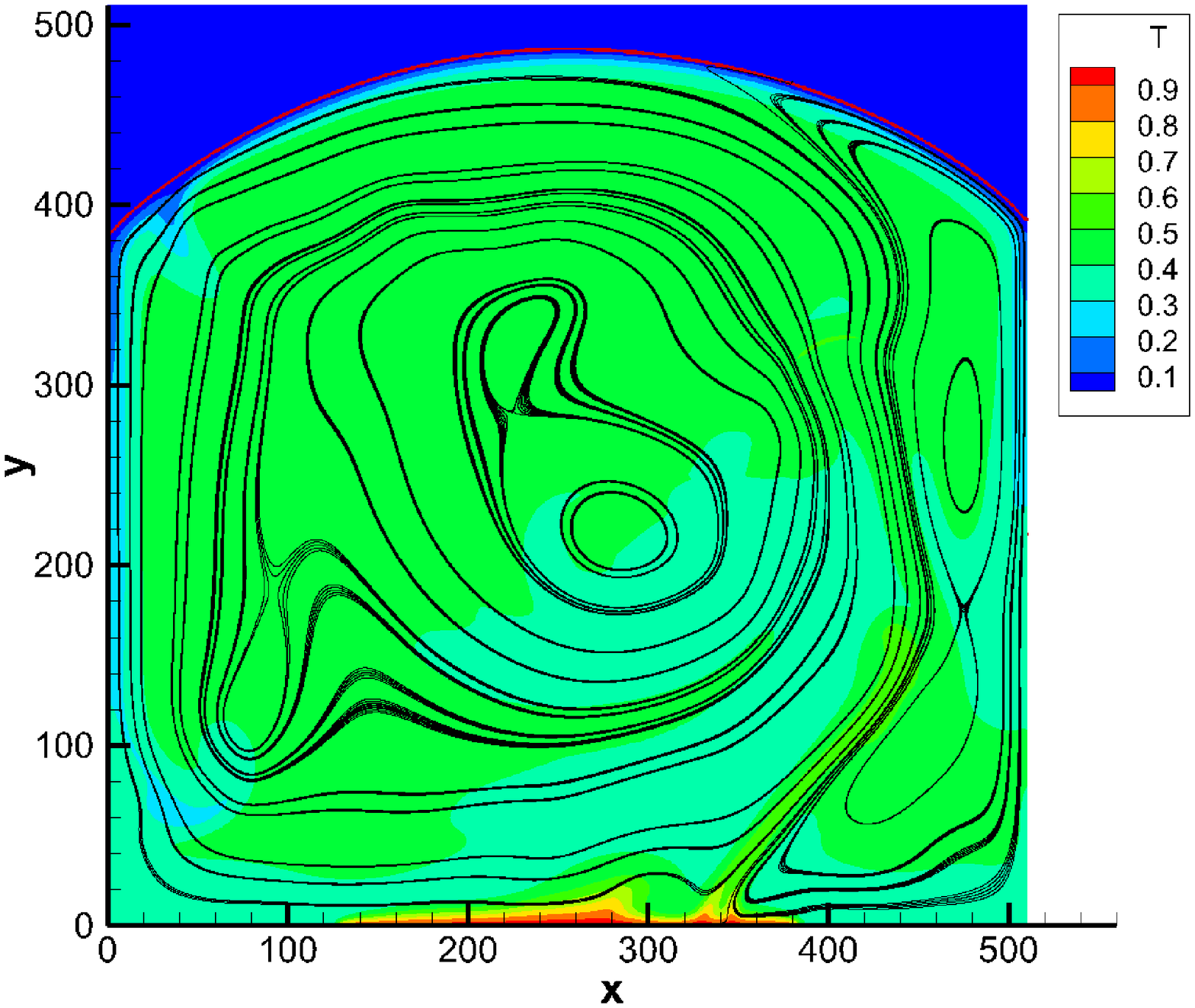}}
\subfigure[]{ \label{fig6:d}
\includegraphics[scale=0.12]{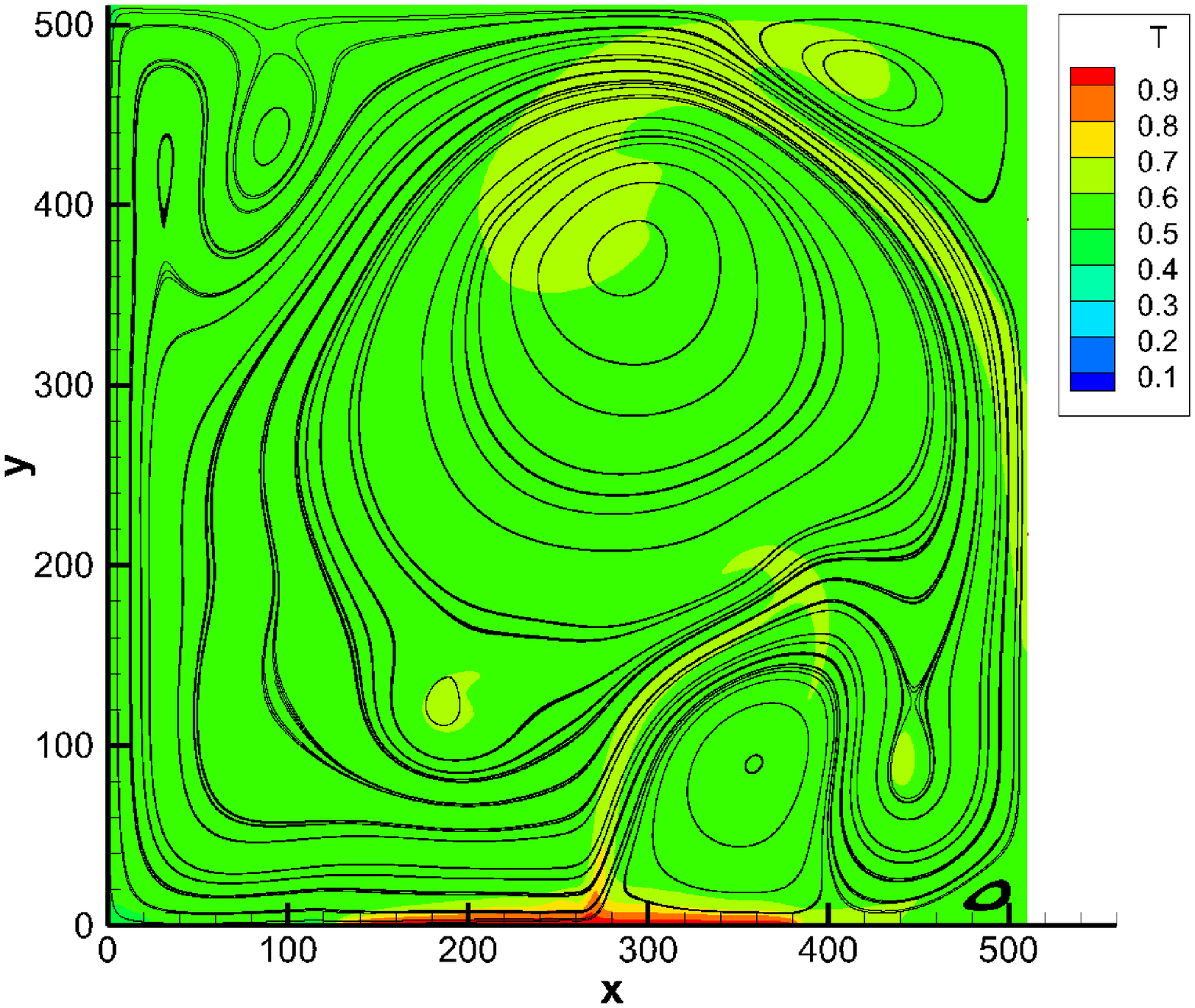}}

\caption{The streamlines of the different cases at different $Fo$ ((a-d): C=0.25L,~(e-h): C=0.375L,~(i-l): C=0.5L;~(a,e,i): $Fo= 0.1$, (b,f,j): $Fo=0.2$, (c,g,k): $Fo=0.3$, (d,h,l): $Fo=0.4$), the phase interface is marked by red line.} \label{fig4}
\end{figure}

In the present investigation, we consider the influence of the C on the melting process, and plot the liquid fraction variation for different cases in the Fig.~\ref{fig3:b} where $T_h=1,~T_m=0,~Ra=10^8,~Pr=6.2,~Ste=0.125,~c=1.0,~\Delta x= 1/512$. As shown in the Fig.~\ref{fig3:b}, when the distance C is changed from 0.25L to 0.5L, the full melting time is decreased. The result indicate that the melting rate can be increased through changing heated region from sides to middle region. The reason can be interpreted by a symmetry argument. As discussed in Ref.~\cite{zhao2018lattice}, the average distance between PCM and heated region is minimized when the heated region is centered, which allows for faster melting. Furthermore, we also we show instantaneous flow and temperature structures in Fig.~\ref{fig4}. From this figures, one can observe that the flows in the melt region are near to be chaotic, and the thermal boundary layers are very thin. In addition, the interface area in the case of C = 0.5L is larger than the interface area in the case of C = 0.25L and C = 0.375L, which results in a larger melting rate at C = 0.5L.

%
%
%
%

\section{Conclusions}
In this work, we present a B-TriRT model to solve the solid-liquid phase change problem, and propose a relational expression to reduce the numerical diffusion across the phase interface, which has been confirmed in Sect.~3.1. In addition we also investigated the the influences of location of the heated region on the melting process at high Rayleigh number, and the results show that the when the heated region is located in the center of bottom wall, the full melting occur earliest. 

\section*{Acknowledgments}

This work is supported by the National Natural Science Foundation of China (Grant No. 51576079) and the National Key Research and Development Program of China (Grant No. 2017YFE0100100).

\section{References}
\bibliographystyle{ws-mplb}
\bibliography{ref}

\end{document}